\documentclass[aip,amsmath,amssymb, reprint, Conference Proceedings]{revtex4-1}

\usepackage{graphicx}% Include figure files
\usepackage{dcolumn}% Align table columns on decimal point
\usepackage{bm}% bold math
\usepackage{inputenc}
\usepackage{graphicx}
\usepackage[T1]{fontenc}
\usepackage{mathptmx}
\usepackage{xcolor,soul}

\begin{document}

\preprint{AIP/123-QED}

\title{Role of symmetry mismatch, charge transfer and low dimensionality in the magnetic order of CaMnO$_3$/SrTiO$_3$ ultrathin heterostructures}
% Force line breaks with \\

\author{M.A. Barral}
\affiliation{Instituto de Nanociencia y Nanotecnolog\'{\i}a (INN CNEA-CONICET), 1650 San Mart\'{\i}n, Argentina}
\affiliation{Departamento de F\'{\i}sica de la Materia Condensada, GIyA-CNEA, Avenida
General Paz 1499, (1650) San Mart\'{\i}n, Pcia. de Buenos Aires, Argentina}
\email{dinapoli@tandar.cnea.gov.ar}

\author{A.M. Llois}
 \affiliation{Instituto de Nanociencia y Nanotecnolog\'{\i}a (INN CNEA-CONICET), 1650 San Mart\'{\i}n, Argentina}
\affiliation{Departamento de F\'{\i}sica de la Materia Condensada, GIyA-CNEA, Avenida
General Paz 1499, (1650) San Mart\'{\i}n, Pcia. de Buenos Aires, Argentina}

\author{S. Di Napoli}
\affiliation{Instituto de Nanociencia y Nanotecnolog\'{\i}a (INN CNEA-CONICET), 1650 San Mart\'{\i}n, Argentina}
\affiliation{Departamento de F\'{\i}sica de la Materia Condensada, GIyA-CNEA, Avenida
General Paz 1499, (1650) San Mart\'{\i}n, Pcia. de Buenos Aires, Argentina}

\date{\today}% It is always \today, today,

\begin{abstract}
By means of \textit{ab initio} calculations, we study the effect of charge transfer and symmetry mismatch on the magnetic and electronic properties of CaMnO$_3$ ultrathin films, epitaxially grown on SrTiO$_3$ (001). We find that the interplay of these degrees of freedom, together with the low dimensionality and the strain imposed by the substrate, changes the bulk occupancy of the Mn $e_g$ orbitals, which determines the magnetic configuration of the ultrathin CMO films. A transition from an insulating G-type to a metallic A-type antiferromagnetic configuration is stabilized.
\end{abstract}

\maketitle

\section{\label{sec:intro} Introduction}
The last decades were characterized by the finding of emergent phenomena in oxide heterostructures, such as two dimensional electron gases at the interface of two insulators~\cite{2004-2DEG} or ferromagnetism between two non-magnetic materials~\cite{Brinkman2007}. These interesting phenomena are a consequence of the strong coupling between spin, orbital, charge and lattice degrees of freedom in this kind of heterostructures. Due to the strong electron-lattice correlations, the electronic and magnetic properties of thin oxide films grown on top of substrates are highly influenced by  strain effects, as the substrate imposes its in-plane lattice constant and symmetry, by charge transfer at the interface and by low dimensionality effects, as at the surface some of the interactions are missing due to the spatial confinement. Another factor that surely influences the film's properties is the symmetry mismatch, which can be tuned by the interfacial oxygen octahedral coupling present at the interface discontinuity.
The thinner the nanostructure, the more important the surface and interface effects due to the increasing surface/volume ratio, opening, in this way, the possibility of an interface engineering. For instance, it was shown in Ref.~\onlinecite{NatMat2016}, that the magnetic and electronic properties of a La$_{2/3}$Sr$_{1/3}$MnO$_3$ (LSMO)  film grown on a NdGaO$_3$ (NGO) substrate could be manipulated by engineering the oxygen network at the unit-cell level. The strong oxygen octahedral coupling  was found to transfer the octahedral rotation of the substrate in the interface region. The first two unit cell layers of LSMO have almost the same tilt angle as NGO and the impact of the octahedral coupling decays rapidly away from the interface and disappears above 4 unit cell layers. Similar effects have been found in LaNiO$_3$/SrTiO$_3$, LaAlO$_3$/SrTiO$_3$ heterostructures~\cite{Triscone2019} and in La$_{2/3}$Sr$_{1/3}$MnO$_3$ grown on SrTiO$_3$~\cite{2014-mismatch}.

In the search after new heterostructured materials suitable for  spintronic devices, consideration of transition metal perovskites is mandatory. Within these perovskites,  CaMnO$_3$ (CMO), which shows a  competition between ferromagnetic double-exchange~\cite{Zener1951,deGennes1960} and antiferromagnetic superexchange interactions~\cite{Goodenough1955}, deserves special attention. The magnetic coupling among Mn ions is very sensitive to the local atomic environment and, therefore, lattice distortions imposed by epitaxial strain or low dimensionality could lead to strong changes in the properties of the thin-films. 

In a previous work \cite{Pedroso2020} we computationally analyzed the effect of high tensile strain and low dimensionality on the magnetic and electronic properties of CaMnO$_3$ ultrathin films imposing the in-plane lattice constants of SrTiO$_3$ (STO), to simulate an epitaxial growth. We found that the combination of both effects yields a change in the magnetic order, from the G-type antiferromagnetic (GAF) structure present in bulk CMO to the A-type antiferromagnetic (AAF) configuration. The GAF structure is characterized by an antiferromagnetic (AFM) coupling between all first neighbors while the AAF one exhibits a ferromagnetic (FM) coupling within the (001) planes and an AFM coupling between adjacent (001) planes. In this case, the FM component of the stabilized magnetic structure is a consequence of self-doping because of charge redistribution originated in the absence of apical oxygens and strain, and not of direct electron doping due to divalent-trivalent chemical substitution or due to the presence of O vacancies, as reported in the past~\cite{La-doped1,La-doped2,Molinari2014}.

In the present work, we analyze the chemical influence  as well as the interface structural rearrangement of the octahedral environment due to the substrate, by performing \textit{ab initio} calculations on CaMnO$_3$ ultrathin films grown on SrTiO$_3$ (001). The  oxygen octahedral distortion is not only a consequence of the strain imposed by the substrate, but also of the discontinuity present at the interface which changes from  $a^0a^0a^0$ (in Glazer's notation) in STO to $a^-a^-c^+$ in CMO, similar to what was found in other heterointerfaces\,\cite{NatMat2016,Triscone2019,tensiones1,tensiones2,tensiones3}. We, thus, consider not only  the symmetry mismatch but also evaluate the charge transfer at the interface of the heterostructure.

\section{\label{sec:compu} Computational details}
We perform \textit{ab initio} calculations within  the framework of Density Functional Theory (DFT) and the projector  augmented wave (PAW) method,~\cite{PAW} as implemented in the Vienna \textit{ab initio} package (VASP)~\cite{VASP,PAW-VASP}. We explicitly treat 10 valence electrons for Ca (3s$^2$3p$^6$4s$^2$), 13 for Mn (3p$^6$3d$^5$4s$^2$), 10 for Sr (3s$^2$3p$^6$4s$^2$), 12 for Ti (3s$^2$3p$^6$4s$
^2$3d$^2$) and 6 for  O (2s$^2$2p$^4$). The local spin density approximation (LSDA) in the parametrization of Ceperley and Alder is used~\cite{LDA1,LDA2}.  All the DFT calculations are performed using a 500~eV energy cutoff in the plane waves basis. As already shown, for manganese perovskites the local spin density approximation successfully predicts the observed stable magnetic phase and the structural parameters~\cite{Picket1999, Picket2000, Nordstrom2017}. We include a Hubbard term with $U=5$~eV and $J=1$~eV within the Lichtenstein implementation~\cite{Liechtenstein95}, for a better treatment of the Mn and Ti $3d$-electrons. The underestimation of the band gap is usual within the local density approximation and, therefore, we set the Hubbard U parameter to the one which better reproduces the experimental band gaps. With these parameters, the relation between the gaps of both materials and the CaMnO$_3$ magnetic ground state are well reproduced~\cite{Nordstrom2017, Piyanzina_2017}.
We approximate the magnetic configuration of the Mn magnetic moments with collinear structures, as  noncollinearity  is known to be quite minimal, giving rise to a small magnetic moment of 0.04$\mu_B$\,\cite{Spaldin2011,Bibes2019}. Therefore, we consider the three relevant antiferromagnetic orders that might be shown by the Mn atoms, being the already defined GAF and AAF as well as the CAF (AFM within the (001) planes and FM between adjacent planes). We also consider the FM structure, for comparison.

For the slab calculations we use a supercell that contains eight layers of CaO alternating  with nine layers of MnO$_2$ grown on top of seven unit cells of SrTiO$_3$ along the (001) direction and a vacuum space of 15\AA~ to avoid the fictitious interaction between slabs due to the periodic boundary conditions. We also consider the TiO$_2$ and the MnO$_2$ terminated surfaces, as it was found that they are the most energetically stable ones~\cite{apl_imbrenda}. Therefore, the interface between both insulators corresponds to SrO/MnO$_2$. To evaluate the integrals within the BZ a 6$\times$6$\times$1 Monkhorst-Pack $k$-point grid is employed and structural relaxations are performed until the forces on each ion are less than 0.01~eV/\AA.

We constrain the first four layers of STO to the LDA optimized bulk structure, and allow the other three layers as well as  the CMO slab to relax their internal coordinates.

\section{\label{sec:results} Results and discussion}

In  a first step, we calculate bulk SrTiO$_3$ and CaMnO$_3$ separately. Bulk computations for STO lead to an optimized lattice constant 
$a_{STO}=3.867$\,\AA,  1\% smaller than the experimental value ($a_{STO}^{exp}=3.905$\,\AA), and a band gap $\Delta_g^{STO} \sim 2.4$\,eV. 
The experimentally determined indirect band gap energy in STO is 3.25 eV, while the direct one is 3.75 eV. The conduction bands are mainly 
composed of Ti 3d $t_{2g}$ and $e_g$ states\cite{2001-STO-exp}. Thin films' band gaps are usually narrower than their bulk counterparts 
due to confinement effects and we obtain a $\Delta_g^{'STO} \sim 1.45$\,eV, for a slab of 7 unit cells. It is worth to mention that these 
results are robust for values of the U parameter in the range 0-5\,eV assigned to the Ti $3d$-orbitals. Regarding CMO, the calculated band 
gap for the bulk stable magnetic configuration, GAF, is $\Delta_g^{CMO} \sim 1.6$\,eV, and it is reduced to $\Delta_g^{'CMO} \sim 0.3$\,eV
for the free standing slab of 8.5 unit cells. As already shown, when the CMO film is tensile strained using the lattice parameters of STO,
the system becomes metallic~\cite{Pedroso2020}.

\begin{figure}[ht!]
\includegraphics[width=1.\columnwidth]{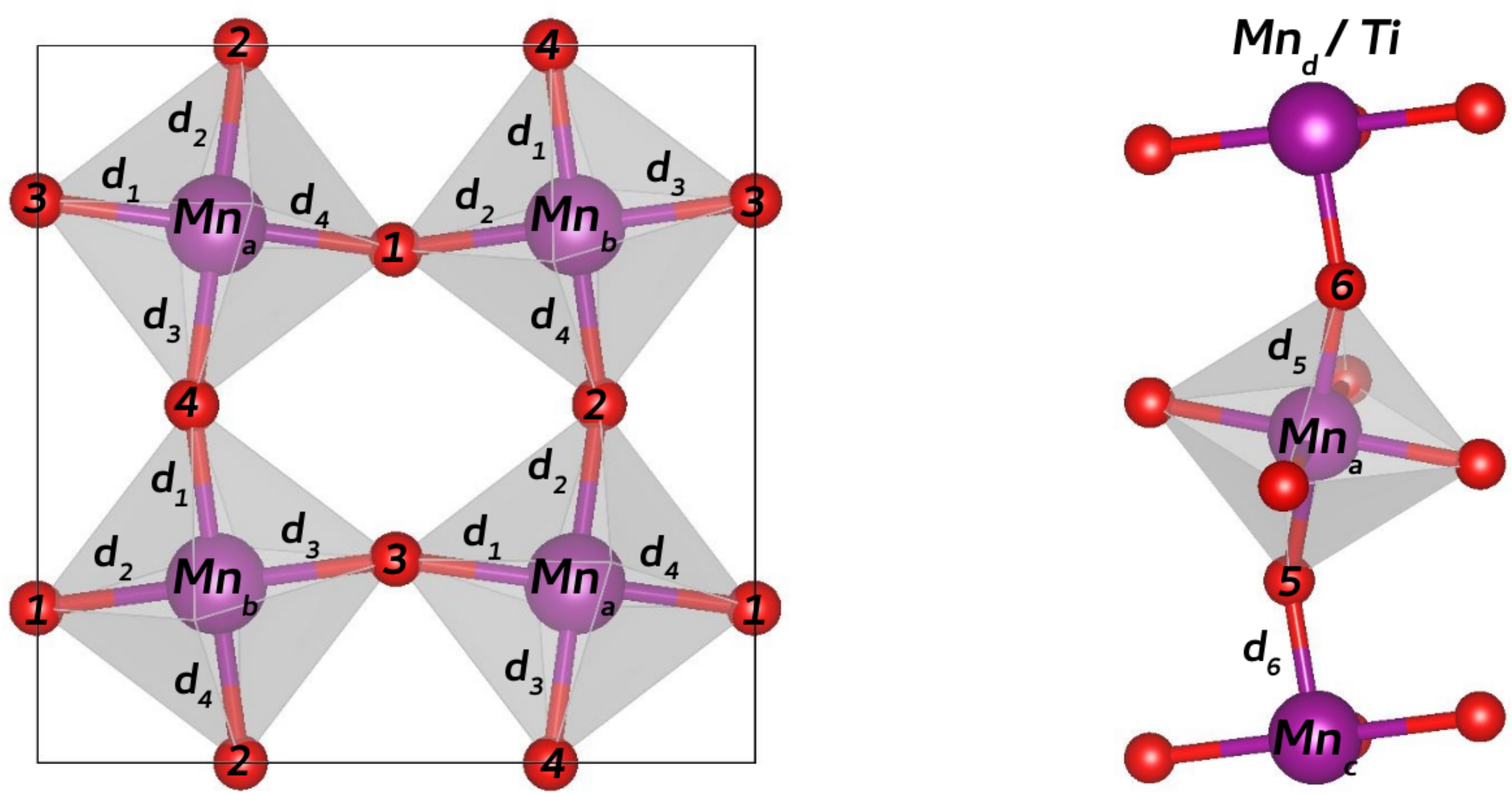}
\caption{(Color online) Top (left) and perpendicular (right) view of a MnO$_2$ surface layer. The defined distances $d_i$ are shown in Table~\ref{tab-structure}.}
\label{fig:structure}
\end{figure}

\begin{table}[ht!]
\centering
\caption{For the AAF magnetic configuration, Mn-O bond lengths (in \AA) and Mn-O-Mn bond angles (in deg) defined in Fig.~\ref{fig:structure}, measured at the different zones of the heterostructure. That is, MnO$_2$ located at the vacuum interface, in the middle layer and at the CMO/STO interface. We also include, for comparison purposes, the corresponding values obtained for the vacumm interface and the middle layer of a free-standing CMO strained-slab.}
\label{tab-structure}
\resizebox{1.\columnwidth}{!}{
\begin{tabular}{|c|c|c|c|c|c|}
\hline
    & \multicolumn{2}{c|}{Free-standing CMO} & \multicolumn{3}{c|}{CMO/STO}\\
    & \multicolumn{2}{c|}{strained-slab} & \multicolumn{3}{c|}{heterostructure}\\
\hline\hline   
    & vacuum    & middle & vacuum    & middle & STO\\
	& interface & layer  & interface & layer  & interface\\
\hline
d$_1$    & 1.79 & 1.97 &   1.79  &    1.97   & 1.92              \\
d$_2$    & 1.80 & 1.97 & 1.81     & 1.97      & 1.94              \\
d$_3$    & 2.00 & 1.98 & 2.00    &  1.98     &  1.96             \\
d$_4$    & 2.29 & 1.98 & 2.28     &   1.98    & 1.99              \\
d$_5$    & 1.78 & 1.85 &  1.77   &   1.84    &  1.86             \\
d$_6$    &  -   & 1.85 & -     & 1.85    & 1.84              \\
$\angle_{Mn_a-O_{1,4}-Mn_b}$ & 151.03 & 156.64 & 151.33 & 156.61 & 162.53 \\
$\angle_{Mn_a-O_{2,3}-Mn_b}$ & 170.49 & 156.66 & 170.89 & 156.49 & 163.30 \\
$\angle_{Mn_a-O_5-Mn_c}$ & 149.43 & 151.32 & 149.37 & 151.34 & 153.41  \\
$\angle_{Mn_a-O_6-Mn_d (Ti)}$ & - & 151.41 & -  & 151.34 & 166.55 \\
\hline				
\end{tabular}
}
\end{table}

In the left panel of Figure\,\ref{fig:structure} we show a schematic top view of a MnO$_2$ surface layer and in the right panel a perpendicular view of a MnO$_6$ octahedron is displayed, where the defined distances $d_i$ and angles are indicated. Both, the tensile strain and the presence of different interfaces  induce changes in the Mn-O bond-lengths and in the  Mn-O-Mn bond angles, the values of all the defined parameters can be found in Table\,\ref{tab-structure}. We separate three zones in the CMO layers of the heterostructure: the MnO$_2$-vacuum interface (named vacuum interface), a MnO$_2$ layer located at the middle of the CMO film and a MnO$_2$ layer in contact with the STO substrate (STO interface). We also include, for comparison, the corresponding values for the free standing CMO strained slab~\cite{Pedroso2020}. 
As it can be seen, the $d_i$ values and Mn$_a$-O$_i$-Mn$_b$ angles in the middle layer are very similar to the ones present in the  middle layer of the free standing strained slab and to those appearing in the strained bulk within the AAF magnetic structure~\cite{Pedroso2020}. This fact indicates that our slab is wide enough to separate the effects of both interfaces. In this middle zone, CaMnO$_3$ adapts its MnO$_6$ octahedral network to the epitaxial strain due to the substrate and the bond-angles increase in the $ab$-plane and decrease along the $c$-axis, while the bond-lengths are larger within the plane and shorter along the perpendicular axis. 

As the film thickness is large enough to decouple the two CMO slab surfaces, the structural reconstruction at the vacuum interface is the same already found for the free standing strained slab, as it can be seen in Table\,\ref{tab-structure}. The absence of the apical oxygens triggers structural distortions around the transition metal atom. The in-plane Mn-Mn bond-lengths equivalency, present in the bulk geometry, is lifted due to the octahedral tilting. Therefore,  two distinct bond-lengths appear and the same happens with the in-plane Mn-O-Mn angles. 

Comparing the distortions at the vacuum interface with the ones induced by the STO interface, we see that the first ones are more important. The bonding distances are, namely, slightly modified by the presence of the STO interface while the bonding angles are changed but not as much as in the vacuum interface. The smaller distortions induced by the STO interface are probably due to the fact that the oxygen atoms located at the interface do not suffer a strong oxygen octahedral rotation when going from the $a^0a^0c^0$ STO to the $a^-a^-c^+$ CMO in the $c$-direction. Clearly, the oxygen octahedral rotation  is much greater at interfaces that have a discontinuity in the octahedral ordering of the type in-phase to out-of-phase, as it happens in La$_{2/3}$Sr$_{1/3}$MnO$_3$/NdGaO$_3$~\cite{NatMat2016}.

It is worth mentioning that the distortions induced by the presence of both interfaces extend within two unit cells away from the discontinuities, before the Mn environment of the strained bulk CMO is recovered.

\begin{figure}[ht!]
\includegraphics[width=1.\columnwidth]{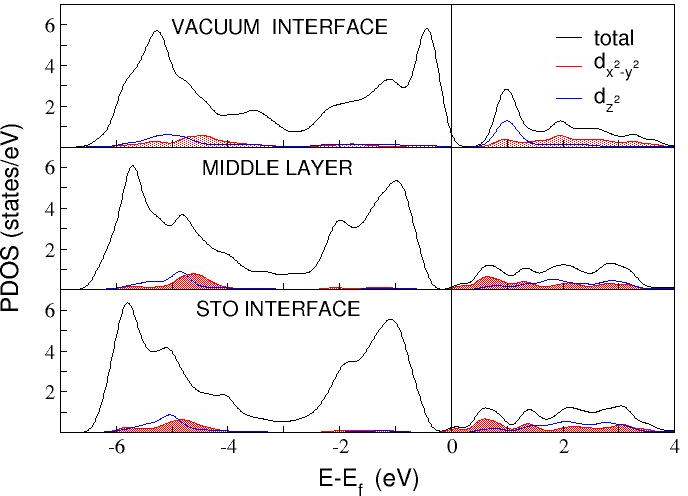}
\caption{(Color online) Spin up partial densities of states (PDOS) of MnO$_2$ layers located at the vacuum interface (top), middle layer (middle) and at the interface with the STO substrate (bottom). Black lines: total DOS of each layer, blue lines: PDOS of each layer projected onto the $d_{z^2}$-symmetry and red lines: PDOS of each layer projected onto the $d_{x^2-y^2}$-symmetry . The occupied $t_{2g}$-orbitals are localized in an  energy range  between -6.5 and -3.5 eV.}
\label{fig:pdos}
\end{figure}

In Figure\,\ref{fig:pdos} we show the total and partial densities of states (PDOS) projected onto the $d_{z^2}$ and $d_{x^2-y^2}$ orbitals, for the MnO$_2$ planes lying in the different regions introduced in Table~\ref{tab-structure}. As expected, tensile strain lifts the degeneracy of the $3d$ $e_g$ orbitals, lowering in energy the $d_{x^2-y^2}$ orbital. 
From the total density of states projected onto the vacuum layer, we can see that the oxygens located at the surface loose part of their charge, and this charge is redistributed among the Mn atoms located in inner layers, starting to occupy the previously empty $d_{x^2-y^2}$ orbitals. On the other hand, charge is also transferred to the $d_{x^2-y^2}$ orbitals of the  Mn atoms located at the STO interface. The electrons self-doping of the $e_g$ states that was already present at the vacuum interface is enhanced by the charge transfer stemming from the STO substrate.

 Both, charge transfer and  self-doping allow the presence of the FM double exchange leading to the stabilization of the AAF magnetic structure, which has more FM pairs than the GAF magnetic configuration of the bulk. The magnetic interaction between the Mn atoms is determined by the competition between AFM superexchange via the Mn $t_{2g}$ electrons and FM double exchange via the Mn $e_g$ electrons. When the $d_{x^2-y^2}$ orbitals are partially occupied the double exchange in the MnO$_2$ planes strengthens the FM ordering while superexchange stabilizes the AFM one between the planes. Within the AAF magnetic structure the CMO film becomes, then, metallic, as it can be also seen from Fig.\,\ref{fig:pdos}. Summarizing, the interplay of epitaxial strain, low dimensionality and charge transfer through the interface changes the occupancy of the $e_g$ orbitals determining the magnetic interactions of the ultrathin CMO films grown on STO substrates.\\

\section{\label{sec:conclu} Conclusions}
In this work we analyze, by means of DFT+U calculations, the  chemical effects at the interface of  CaMnO$_3$ ultrathin films epitaxially grown on top of SrTiO$_3$ (001) substrates. We address
the role of lattice mismatch and charge transfer on the magnetic and electronic properties of these thin films. 

From the structural analysis we find that in this heterostructure, the octahedral rotation pattern of strained CMO bulk is preserved along the film and that the octahedral distortions  are important mainly in the vicinity of the vacuum and STO interfaces, characterized by a MnO$_6$ octahedral discontinuity.

Concerning the magnetic properties, we obtain that the AAF magnetic configuration with in-plane FM pairs is more stable than the GAF configuration present in bulk CMO. This magnetic transition is driven by the presence of self-doping electrons at the vacuum surface, as already found in a previous work (Ref.~\onlinecite{Pedroso2020}), which is strengthened by the charge transferred by the STO substrate to the interface. This charge transfer amplifies the charge redistribution driven by tensile strain and quantum confinement. Interestingly, coupled by the magnetic transition from GAF to AAF, we find an insulator to  metal transition in the CMO film. \\

\section*{Acknowledgments}
This work was partially supported by PICT-2016-0867 of the ANPCyT, Argentina, and by H2020-MSCA-RISE-2016 SPICOLOST Project Nº 734187.  

\section*{Data Availability}
The data that support the findings of this study are available from the corresponding author upon reasonable request.

\bibliography{biblio}
\end{document}